\documentclass{aa}  
\usepackage{mathabx}
\usepackage{hyperref}
\usepackage{txfonts}

\begin{document} 

\title{The high-energy radiation environment of the habitable-zone super-Earth LHS~1140b}
%   \subtitle{UV and X-ray characterization}
\titlerunning{The high-energy radiation environment of LHS~1140b}
\authorrunning{R.~Spinelli et al.}
   \author{R.~Spinelli \thanks{email: r.spinelli5@campus.unimib.it}\inst{1}, F.~Borsa\inst{2}, G.~Ghirlanda\inst{1,2},
   G.~Ghisellini\inst{2}, S.~Campana\inst{2}, F.~Haardt\inst{3,4}, E.~Poretti\inst{2,5}
         % C. Ptolemy\inst{1}\fnmsep\thanks{Just to show the usage
         % of the elements in the author field}
          }
   \institute{Dipartimento di Fisica G. Occhialini, Universit{\`a} Milano--Bicocca, Piazza della Scienza~3, 20126 Milano, Italy
%\email{c.ptolemy@hipparch.uheaven.space}
%\thanks{The university of heaven temporarily does not
    %       accept e-mails}
 %\email{r.spinelli5@campus.unimib.it}
    \and  
   INAF -- Osservatorio Astronomico di Brera, Via E. Bianchi 46, 23807 Merate (LC), Italy
       \and  
 Dipartimento di Scienza e Alta Tecnologia, Universit\`a dell'Insubria, Via Valleggio 11, 22100 Como, Italy
 \and
 INFN -- Sezione Milano--Bicocca, Piazza della Scienza~3, 20126 Milano, Italy
       \and  
 Fundaci{\'o}n Galileo Galilei -- INAF, Rambla Jos{\'e} Ana Fernandez P{\'e}rez 7, 38712 Bre$\tilde{\rm n}$a Baja, TF, Spain
             }

   \date{Received March 26, 2019; accepted }

% \abstract{}{}{}{}{} 
% 5 {} token are mandatory
 
  \abstract
   {In the last few years many exoplanets in the habitable zone (HZ) of M-dwarfs have been discovered, but the X-ray/UV activity of cool stars is very different from that of our Sun. The high-energy radiation environment influences the habitability, plays a crucial role for abiogenesis, and impacts the chemistry and evolution of planetary atmospheres. LHS~1140b is one of the most interesting exoplanets discovered. It is a super-Earth-size planet orbiting in the HZ of LHS~1140, an M4.5 dwarf at $\sim$15 parsecs.}
   {In this work, we present the results of the analysis of a {\it Swift} X-ray/UV observing campaign. We characterize for the first time the X-ray/UV radiation environment of LHS~1140b.}
   {We measure the variability of the near ultraviolet (NUV) flux and estimate the far ultraviolet (FUV) flux with a correlation between FUV$_{1344-1786\AA}$ and NUV$_{1771-2831\AA}$ flux obtained using the  sample of low-mass stars in the GALEX archive. We highlight the presence of a dominating X-ray source close to the J2000 coordinates of LHS~1140, characterize its spectrum, and derive an X-ray flux upper limit for LHS~1140. 
   We find that this contaminant source could have influenced the previously estimated spectral energy distribution.
   }
   {No significant variation of the NUV$_{1771-2831\AA}$ flux of LHS~1140 is found over 3 months, and we do not observe any flare during the 38~ks on the target. LHS~1140 is in the 25th percentile of least variable M4-M5 dwarfs of the GALEX sample. 
   Analyzing the UV flux experienced by the HZ planet LHS~1140b, we find that outside the atmosphere it receives a NUV$_{1771-2831\AA}$ flux $<2\%$ with respect to that of the present-day Earth, while the FUV$_{1344-1786\AA}$/NUV$_{1771-2831\AA}$ ratio is $\sim$100-200 times higher. This represents a lower limit to the true FUV/NUV ratio since the FUV$_{1344-1786\AA}$ band does not include Lyman-alpha,    which dominates the FUV output of low-mass stars. This is a warning for future searches for biomarkers, which must take into account this high ratio.}
   {The relatively low level and stability of UV flux experienced by LHS~1140b should be favorable for its present-day habitability.}


   \maketitle
%
%________________________________________________________________

\section{Introduction}
In the current state of exoplanet exploration, extrasolar planets orbiting M-dwarfs represent the most promising targets to discover habitable worlds \citep[e.g.,][]{2016PhR...663....1S}. Besides being the most abundant stellar population in the Galaxy \citep[$\sim$75\%,][]{2010AJ....139.2679B}, M-dwarfs facilitate the discovery of planets with methods using transits and radial velocities \citep{2008PASP..120..317N, 2014IAUS..299..395Q}. Moreover, given their low masses, M-dwarfs are extremely long-lived and offer long timescales for possible biological evolution on their orbiting planets \citep{2016PhR...663....1S}. 
The habitable zone (HZ) is typically defined as the orbital distance at which the steady stellar irradiation leads to a suitable temperature for the presence of liquid water on the planetary surface \citep[e.g.,][]{1993Icar..101..108K,2013ApJ...765..131K}. Recent exoplanet population studies suggest that planets orbiting M-dwarfs are common: in particular, \citet{2015ApJ...807...45D} estimated that on average, for every
seven M-dwarfs, there is at least one Earth-size planet orbiting in the HZ. 
The HZ of M-dwarfs is closer ($\sim$0.1 AU) to the star with respect to earlier spectral types, and they are also known to be very strong X-ray/UV emitters as a consequence of magnetic activity in their outer atmospheres. This implies a possible higher exposure of HZ exoplanets orbiting M-dwarfs to a variable and energetic environment. \citet{2013ApJ...763..149F} highlighted the possibility that the high-energy variability of M-dwarfs occurs with strong flare behaviors, even in M-dwarfs previously considered to be weakly active (see also \citealt{2018ApJ...867...71L}). This high-energy environment has several implications for the evolution of the HZ planets orbiting M-dwarfs and for the possible emergence of life. 

X-ray coronal emission is one of the main drivers of planetary atmospheric erosion \citep{2010A&A...511L...8S}. Ultraviolet radiation can inhibit photosynthesis, induce DNA destruction, cause damage to various species of proteins and lipids \citep{2007Icar..192..582B}, and destroy nascent biomolecules \citep{Sag}. On the other hand, as suggested by recent experimental and theoretical studies \citep{2017ApJ...843..110R}, UV radiation is a crucial element for prebiotic photochemistry, and  especially for the synthesis of RNA, which favors the emergence of life. In addition, the far UV (FUV, 912--1700 $\AA$) emission of M-dwarfs is usually comparable with the solar one, while the near UV (NUV, 1700--3200 $\AA$) flux of M-dwarfs seen by an exoplanet in the habitable zone can be up to three orders of magnitude fainter than the solar radiation received at Earth \citep{2012ApJ...750L..32F, 2013ApJ...763..149F}. This kind of emission has several implications for the atmospheric composition of planets and can lead to the formation of false-positive biosignatures \citep{2014E&PSL.385...22T}. 
Indeed, as suggested by \citet{2015ApJ...812..137H}, the typical FUV/NUV emission of M-dwarfs can induce the abiotic production of $\rm O_2$ through photolysis of $\rm CO_2$ followed by recombination of $\rm O$ atoms. This is only one of the numerous abiotic oxygen-formation pathways hypothesized in the literature. For example, abiotic $\rm O_2$ can also be produced in HZ planets by the runaway greenhouse effect after  planet formation  due to the superluminous pre-main sequence phase of M-dwarfs (\citealt{2015AsBio..15..119L}), or in atmospheres with low amounts of noncondensing gases without a cold trap (\citealt{2014ApJ...785L..20W}; see also \citealt{2018AsBio..18..663S} for a recent review). For these reasons, the characterization of the emission of M-dwarfs in UV bands is crucial to constraining the possibility of life as we know it forming and surviving in their HZs.
In this work we use the GALEX far UV ($\rm FUV_{G}$, 1344--1786 {$\AA$}) and the GALEX near UV ({$\rm NUV_{G}$}, 1771--2831 {$\AA$}) bandpasses to compare LHS~1140 with a large sample of M-dwarfs. We note that the GALEX bandpasses do not include important spectral tracers (e.g., Lyman-$\rm \alpha$, Mg II, C II, NV) and therefore only encompass part of the astrobiologically important UV irradiance band.

\object{LHS~1140} is among the most interesting systems discovered up to now. The parent star is a M4.5 dwarf ($\rm 0.179\, M_{\Sun}$, $\rm 0.2139\, R_{\Sun}$) with a rotational period of $\sim$131 days \citep{2017Natur.544..333D,2018arXiv180709365N}. It hosts \object{LHS~1140b} \citep{2017Natur.544..333D}, a super-Earth-size planet in the HZ that transits the host star every $\sim$24.7 days, for which both radius and mass could be measured \citep{2017Natur.544..333D}. Recently, a further two planets were proposed to be hosted in the system, with periods of $\sim$3.8 and $\sim$90 days, with a more in-depth analysis of radial-velocity data \citep{2018arXiv180702483F}. Independently, the shorter-period planet was also found to transit its host \citep{ment}. The properties of the host star and its planets were recomputed \citep{2018ApJ...861L..21K} after the new Gaia DR2 \citep{2018A&A...616A...1G} parallax ($\rm 14.993 \pm 0.015 \ pc$). While LHS~1140b still orbits in the HZ (but with more incident flux from the star), its newly estimated radius is larger by $\sim$20\% ($\rm 1.73 R_{\Earth}$) with a consequent strong reduction of the planet density and a higher suitability for transmission spectroscopy studies. The day-side equilibrium temperature estimated by \citet{ment}, assuming a zero Bond albedo, is $\rm 235\,K$ which would imply that the planet is in a snowball state. However, as \citet{2018arXiv181206606D} recall, the equilibrium temperature does not take into account the atmospheric greenhouse effect that increases the surface temperature of planets and, in the case of LHS~1140b, the chances of habitability. 
We note that, given the small distance from its parent star, LHS1140b is possibly tidally locked \citep{2017CeMDA.129..509B}.
The LHS~1140 system represents a primary target for present and future exoplanet research. In view of the above, in order to characterize the possibility of life (as we know it) emerging on the HZ planet LHS1140b, it is important to characterize its high-energy environment. We consider the high-energy FUV, NUV, and XUV (5-911 ${\AA}$) radiation bands \citep{2016ApJ...820...89F}.
In this work, we present  the results of the analysis of a 38--ks {\it Swift} X-ray/UV observing campaign (Cycle 14; Sect.~\ref{sec:obs}). We derive an upper limit on the stellar X-ray emission (Sect.~\ref{sec:results}), and characterize for the first time UV properties and variability (Sect.~\ref{sec:conclu}). 
Our results are discussed in the context of UV illumination for the exoplanet LHS~1140b.  

\section{Observations and data analysis\label{sec:obs}}
Observations of LHS~1140 were performed with the Neil Gehrels {\it Swift} Observatory \citep{2004ApJ...611.1005G} through a dedicated project (Principal Investigator: F. Borsa). {\it Swift} obtains X-ray and UV data simultaneously thanks to the co-aligned XRT and UVOT. Seventeen observations, none of which were performed during planetary transits, are distributed between December 2017 and February 2018. Table~\ref{tab1140} lists the observation log. 

\subsection{{\it Swift} UVOT}

\begin{figure*}
   \centering
   \includegraphics[scale=0.28]{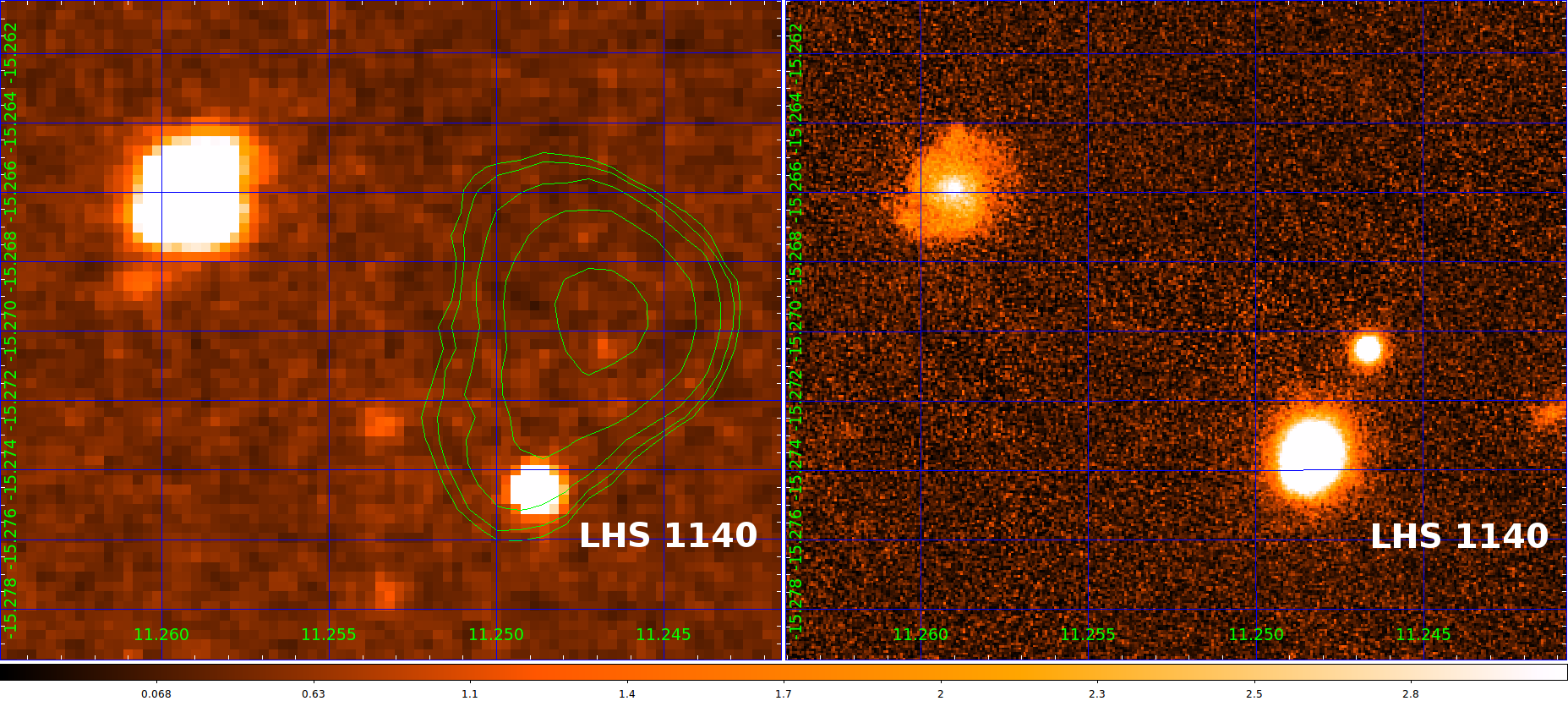}
   \caption{{(\it Left panel)}: {\it Swift} UVOT UVW2 filter image (color) with overlay contour levels of the XRT image (green lines). The XRT centroid coincides with a background source present in the PanSTARRS archive. {(\it Right panel)}: PanSTARRS image (g band) obtained with observations between 2009 and 2014. The coordinates of LHS 1140 do not coincide in these two images due to its proper motion. The object in the upper left is LEDA 913494, a galaxy present in the  Hyperleda database \citep{2003A&A...412...45P}.}    
   \label{im_UVvsX}%
    \end{figure*}

UVOT images (taken with the UVW2 filter) were analyzed with the public software \texttt{HEASOFT (v6.25)} package. The most recent version of the calibration database was used. All exposures within each observation were summed together. Photometry
was performed within a circular source-extraction region of 5 arcsec in radius. The background was extracted from a circular region with a radius of about 20 arcsec, close to our target but without contamination from other sources. A bright UV source (Fig.~\ref{im_UVvsX}) is detected at RA=00h 44m 59.6s  and Dec=$-$15$^{\circ}$ 16$'$ 28.5$''$ (Fig.~\ref{im_UVvsX}). This position is $\sim$12 arcsec off the J2000 coordinates of LHS 1140: this offset is consistent with the proper motion of the source\footnote{\url{http://simbad.u-strasbg.fr/simbad/sim-id?Ident=LHS+1140}}, that is, 317.585$\pm$0.121 mas yr$^{-1}$ in right ascension and  -596.617$\pm$0.085 mas yr$^{-1}$ in declination. The position of LHS~1140 in our UV image matches the one reported in the Gaia Archive\footnote{ \url{https://gea.esac.esa.int/archive/}}.
%______________________________________________ Gamma_1 (lg rho, lg e)
%----------------------------------------------------------------
The source was detected with a significance of between $3.3 \sigma$ and $19.7 \sigma$ in  the 17 epochs. 
The magnitude and flux density of the individual epochs (as reported in Table~\ref{tab1140}) were computed with the tool \texttt{uvotsource}. We also stacked all the observations together with the tool \texttt{uvotimsum} to obtain an average flux density and produce the image shown in Fig.~\ref{im_UVvsX}. 
\subsection{{\it Swift} XRT\label{sec:xrt}}
All {\it Swift} XRT observations were combined in a single data set of $\sim$38 ksec through the public service of the University of Leicester\footnote{\url{http://www.swift.ac.uk/user_objects/}} and the image and spectrum accumulated over this time interval were retrieved. The analysis of the X-ray image was performed with \texttt{Ximage (v4.5.1)}. A source close to the J2000 coordinates of LHS~1140 is detected with a $\rm 3.2 \sigma$ significance. This was obtained with the \texttt{sosta} tool of \texttt{Ximage} by selecting a box region with a half radius of 18.2 pixels (centered on the source). This source has a count rate of $(5.44\pm1.70)\times10^{-4}$ counts s$^{-1}$. In Fig.~\ref{im_UVvsX} the X-ray contours are shown (green lines) overlayed on the UV image. The centroid of the X-ray emission does not coincide with the position (evolved at the epoch of our observations for the high proper motion) of LHS~1140 as detected in the simultaneous UVOT observations. The separation between LHS~1140 and the centroid of the X-ray emission is $\sim$11 arcsec. However, the X-ray contours show some extension towards the present position of LHS~1140 (see Fig.~\ref{im_UVvsX}). By searching in the PanSTARRS archive (\url{https://catalogs.mast.stsci.edu/}) we found that the possible X-ray emission centroid is coincident with a background optical source (it is unlikely to be a foreground source given the distance of our target). 

We further investigated the properties of this contaminating X-ray source by exploiting our XRT observation. Data were extracted from a circular region with a radius of 20 pixels. Background data were extracted from a  larger region close-by, devoid of sources. Source photons were then binned to have 20 counts per spectral bin. The spectrum was analyzed with the \texttt{Xspec (v12.9)}. 
The low signal does not allow for a more refined characterization than a single power law (with an absorption column density fixed to the value $\rm N_{H}=5\times 10^{18}$ $\rm {cm^{-2}}$, in agreement with LIC model of \citealt{2000ApJ...534..825R, 2008ApJ...673..283R}\footnote{\url{http://lism.wesleyan.edu/ColoradoLIC.html}}). The power-law photon spectral index is  $\Gamma=1.78^{+1.07}_{-0.92}$ (90\% confidence error interval) and the reduced $\tilde{\chi}^2=0.8$ (13 degrees of freedom). The unabsorbed source flux (integrated in the 0.3-10 keV energy range) is $1.42^{+0.66}_{-0.45} \times 10^{-14}$ erg cm$^{-2}$ s$^{-1}$.

Given the elongated shape of the X-ray emission as observed by XRT, we estimated a 3$\sigma$ upper limit on the possible X-ray flux at the position of LHS~1140 by masking the counts of the dominating source. We derive a count rate upper limit of 4.95$\times$10$^{-4}$ counts s$^{-1}$ at the current position of LHS~1140. Assuming the spectrum of the contaminating X-ray source, we estimate a $3\sigma$ upper limit on the flux (integrated in the 0.3-10 keV energy range) of ${\sim 1.29\times 10^{-14}}$ erg cm$^{-2}$ s$^{-1}$ and a flux upper limit of $\sim 3.20\times 10^{-15}$ erg cm$^{-2}$ s$^{-1}$ keV$^{-1}$ at 1 keV. We estimate an X-ray luminosity (0.1-3 keV) of $\sim 3 \times 10^{26}$ erg s$^{-1}$ which, given the uncertainties, is consistent with typical values observed in M-dwarfs \citep[e.g.,][]{2016ApJ...821...81G, 2017MNRAS.465L..74W}.

%__________________________________________________ One column table
\begin{table*}
\centering
\begin{tabular}{ccccc}
\hline
  \multicolumn{1}{c}{Observation id} &
  \multicolumn{1}{c}{Date} &
 \multicolumn{1}{c}{Exposure time [s]}  &
  \multicolumn{1}{c}{mag(AB)} &
  \multicolumn{1}{c}{Flux density ($\rm 2032 \AA$) [mJy]}\\
\hline
87541001 & 2017 Nov. 08 & 393.18 & 21.65$\pm$0.30 & 0.0079$\pm$0.0022  \\
87541002 & 2017 Nov. 09 & 757.82 &21.20$\pm$0.16 & 0.0120$\pm$0.0017  \\
87541003 & 2017 Nov. 11 & 207.83 &21.31$\pm$0.33 & 0.0109$\pm$0.0033 \\
87541004 & 2017 Dec. 14 & 317.92 &21.34$\pm$0.27 & 0.0105$\pm$0.0026 \\
87541005 & 2017 Dec. 22 & 284.56 &21.08$\pm$0.25 & 0.0135$\pm$0.0031 \\
87541006 & 2017 Dec. 24 & 2894.13 &21.13$\pm$0.08 & 0.0128$\pm$0.0010 \\
87541007 & 2017 Dec. 30 & 2316.91 &21.15$\pm$0.10 & 0.0126$\pm$0.0011 \\
87541008 & 2018 Jan. 03 & 1629.47 &21.12$\pm$0.11 & 0.0129$\pm$0.0013 \\
87541009 & 2018 Jan. 19 & 1517.95 &21.12$\pm$0.11 & 0.0130$\pm$0.0014  \\
87541010 & 2018 Jan. 26 & 6843.58 &21.14$\pm$0.06 & 0.0127$\pm$0.0003  \\
87541011 & 2018 Jan. 29 & 6869.48 &21.15$\pm$0.06 & 0.0126$\pm$0.0007  \\
87541012 & 2018 Jan. 31 & 5143.95 &21.18$\pm$0.07 & 0.0122$\pm$0.0008 \\
87541013 & 2018 Feb. 01 & 1696.71&21.02$\pm$0.10 & 0.0142$\pm$0.0014  \\
87541014 & 2018 Feb. 02 & 1440.95&21.27$\pm$0.14 & 0.0112$\pm$0.0014  \\
87541015 & 2018 Feb. 04 & 1096.43 &21.13$\pm$0.15 & 0.0128$\pm$0.0018  \\
87541016 & 2018 Feb. 06 & 1408.11&21.19$\pm$0.14 & 0.0121$\pm$0.0015  \\
87541017 & 2018 Feb. 08 &2549.22 &21.01$\pm$0.09 & 0.0143$\pm$0.0012  \\
\hline\end{tabular}\caption{UVOT observations of LHS~1140 with  UVW2 filter.}
\label{tab1140}
\end{table*}
%
%                                     Two column figure (place early!)
%______________________________________________ Gamma_1 (lg rho, lg e)
   \begin{figure}
   \centering
   \includegraphics[scale=0.4]{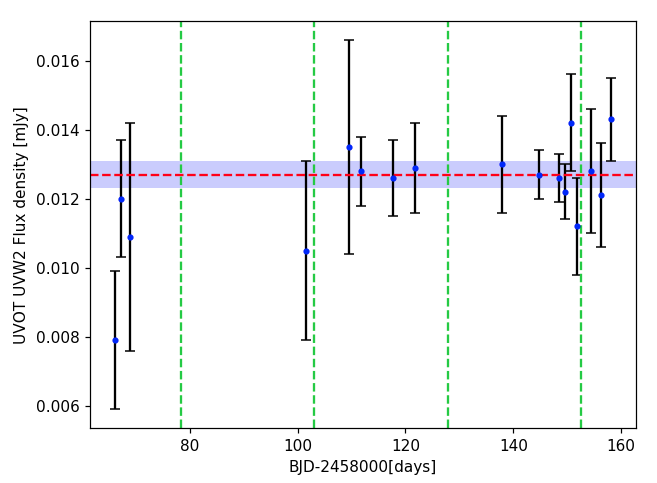}
   \caption{LHS 1140 UVOT-UVW2 flux-density light curve. Error bars represent the 1$\sigma$ uncertainty. The vertical green dashed lines are the epochs of transit of LHS~1140b. The horizontal dashed red line is the average flux density ($\rm 12.7 \ \mu Jy$) as obtained from the analysis of the stacked UV image (Fig.~\ref{im_UVvsX}) with its 1$\sigma$ uncertainty ($\rm 0.4 \ \mu Jy$, shaded region).} 
              \label{FigGam}%
    \end{figure}
%----------------------------------------------------------------

%______________________________________________ Gamma_1 (lg rho, lg e)
   \begin{figure}
   \centering
   \includegraphics[scale=0.4]{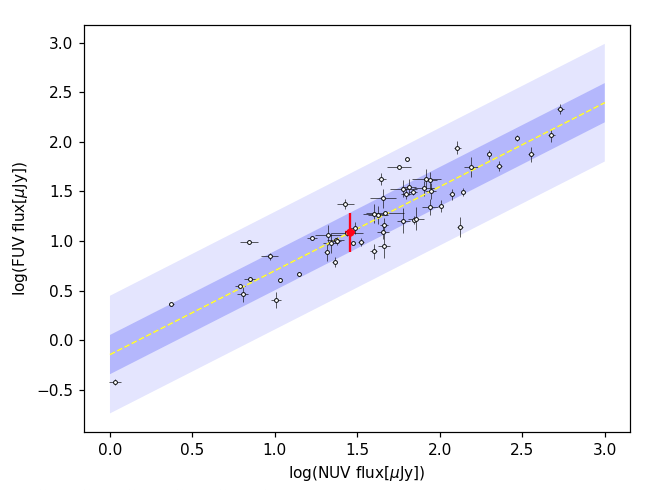}
   \caption{Correlation between the $\rm FUV_G$ and the $\rm {NUV_G}$ flux density of M4-M5 stars. The yellow dashed line is the least-squares fit of the data points. The red point shows the position of LHS 1140 as obtained through our measure of the $\rm NUV_G$ flux density (Sect.~2). The shaded regions mark the $1\sigma$ and $3\sigma$ dispersion (measured perpendicular to the best fit line) of the data points.} 
              \label{im_correlation}%
    \end{figure}
%----------------------------------------------------------------

\section{Results\label{sec:results}}

Figure~\ref{FigGam} shows the flux-density light curve of LHS~1140 as obtained from the 17 UVOT observations (Table~\ref{tab1140}). 
Our UV flux observations, being centered at 2032 $\AA$, are in the $\rm NUV_G$ band.
No significant variations of the UV flux of LHS~1140 are found over the 17 observations distributed over almost 3 months. 
A Generalized Lomb Scargle analysis \citep{2009A&A...496..577Z} does not identify any significant periodicity in our UV flux light curve. Unfortunately, our time coverage ($\sim$90 d) is
shorter than the rotational period ($\sim$131 d): this hampers detection of a possible rotational modulation.       
As an alternative method to quantify the UV variability, we used the $\rm MAD_{rel}$ index \citep{2017AJ....154...67M}. We applied this method to LHS~1140, estimating the median absolute deviation divided by the median of the $\rm NUV_G$ flux. To account for the errors on the flux we generated $10^4$ samples of 17 observations with a Monte Carlo sampling. The mean and standard deviation of the $\rm MAD_{rel}$ of these samples are $0.09 \pm 0.03$. 

\citet{2017AJ....154...67M} used simultaneous $\rm NUV_G$ and $\rm FUV_G$  observations of 145 low-mass stars taken from the  sample of low-mass stars with multiple GALEX ultraviolet observations (GUV) in the GALEX archive\footnote{\url{http://vizier.cfa.harvard.edu/viz-bin/VizieR?-source=J/AJ/154/67}} to find the correlation between $\rm NUV_G$ and $\rm FUV_G$ emission for each spectral type. The NUV and FUV GALEX data were also used by \cite{2013MNRAS.431.2063S} to study M-dwarfs activity.  To quantify the variability of the $\rm NUV_G$ flux of LHS~1140 we compared its $\rm MAD_{rel}$  with a sample of 98 M-dwarfs (from M4 to M5 spectral type: 38 M4, 6 M4.1, 1 M4.2, 3 M4.3, 31 M4.5, 1 M4.8, 2 M4.9, 16 M5) selected from the GUV sample. Our value for LHS~1140 is the 25th percentile for the $\rm MAD_{rel}$ distribution of the sample. LHS~1140 is in the group of least variable M4-M5 dwarfs.

To estimate the $\rm FUV_G$ flux of LHS~1140 from the $\rm NUV_G$ flux observed by UVOT, we selected 57 M-dwarfs (of spectral type from M4 to M5) with simultaneous $\rm NUV_G$ and $\rm FUV_G$ observations from the GUV sample. From this sample, a source with an evident $\rm FUV_G$ flare \citep[as noted also in][]{2017AJ....154...67M} was removed. The $\rm NUV_G$ and $\rm FUV_G$ fluxes in the GALEX archive are already scaled to a distance of 10 pc. These data are shown in Fig. \ref{im_correlation}. A least-squares fit produces the following log-correlation: 
\begin{equation}
 Log(f_{\rm FUV_G})=0.847(\pm 0.05) \times Log(f_{\rm NUV_G}) -0.145 (\pm 0.09)
 \label{eq:1}
.\end{equation}
For LHS~1140, the $\rm NUV_G$ flux scaled to 10 pc is $\rm 28.6 \pm 0.9\ \mu Jy$. The scatter of the data points computed perpendicular with respect to the best-fit correlation was modeled with a Gaussian distribution. 
Using Eq.\ref{eq:1} we estimated for LHS~1140 a $\rm FUV_G$ flux scaled to 10 pc of $\rm 12.2 \pm 2.4 \ \mu Jy$. We measure a NUV magnitude (AB system) of 21.1, while our estimated FUV magnitude is 22.1. These values are consistent with the nondetection of our target in the GALEX AIS (All Sky Imaging Survey) images. In fact, the GALEX AIS data reach limit  magnitudes (5$\sigma$) of 20.8 for the NUV and 19.9 for the FUV \citep{2007ApJS..173..682M}.

In addition, in Sect. \ref{sec:xrt} we report the discovery of a contaminating X-ray source that was close to LHS~1140 around the year 2000 but is no longer like that thanks to the high proper motion of LHS 1140. We therefore visually inspected the images available in the literature (from CDS Portal: \url{https://cds.u-strasbg.fr/}) and reported in Table~\ref{tab1140_sed}, to check if the two sources (LHS~1140 and the contaminant one) were superposed. In the end we decided to exclude the 2MASS data, which were collected between 1997 June and 2001 February because in the corresponding images LHS~1140 and the background source were spatially coincident. Figure 4 shows a  plot of the spectral energy distribution (SED) of the star and  the black-body temperature recalculated using only fluxes obtained from optical and IR images where the two sources are resolved. We did not use the UV fluxes because of the known chromospheric M-dwarfs emission \citep[e.g.,][]{2013ApJ...763..149F, 2016MNRAS.463.1844S}. We obtain a new estimate of effective temperature of LHS~1140 $T_{\rm eff}=\rm 3016 \pm 29 \ K$, which is $\sim$200 K lower than previous estimates.

%______________________________________________ Gamma_1 (lg rho, lg e)
   \begin{figure}
   \centering
   \includegraphics[scale=0.4]{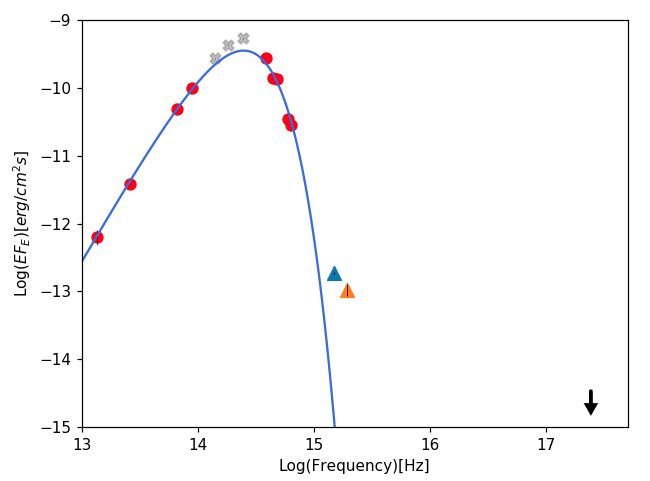}
   \caption{LHS~1140 $EF_{\rm E}$ SED, calculated with data from the Centre de Donn{\'e}es astronomiques de Strasbourg (CDS Portal: \url{https://cds.u-strasbg.fr/}). The points are listed in Table~\ref{tab1140_sed}. Red points mark the black-body temperature estimation. Gray symbols are 2MASS data. UVOT UVW2 NUV flux is represented by the blue triangle. $\rm FUV_G$ flux estimated with $\rm FUV_G-NUV_G$ correlation is represented by the orange triangle. The black arrow shows the 3$\sigma$ upper limit to the X-ray flux of LHS~1140 at 1 keV. The blue line is the black body model with a temperature of 3015 K.} 
              \label{im:sed}%
    \end{figure}
%----------------------------------------------------------------

\begin{table}
\centering
\begin{tabular}{cll}
\hline
  \multicolumn{1}{c}{Filter}&
  \multicolumn{1}{c}{Flux [Jy]} &
  \multicolumn{1}{c}{Flux err [Jy]}  \\
\hline
Gaia:G & 0.0318 & 0.0001        \\
  GAIA/GAIA2:Grp & 0.0722   &  0.0001     \\
GAIA/GAIA2:G  &  0.0282   &  0.0001      \\  
  GAIA/GAIA2:Gbp &  0.00595  &  0.00002            \\
  WISE:W4  &  0.0046 &   0.0011            \\
  WISE:W3 &  0.0147  &   0.0003    \\
   WISE:W2 &  0.0751  &   0.0013                   \\
WISE:W1  &  0.111   &   0.002     \\
 PAN-STARRS/PS1:g &  0.00459 &   0.00002          \\
2MASS:Ks &  0.2     &   0.004              \\
 2MASS:H &  0.242   &   0.006              \\
 2MASS:J&  0.225   &   0.005               \\
\hline\end{tabular}\caption{Optical and IR LHS1140 flux measurements used in Fig.~\ref{im:sed} (CDS Portal \url{http://cdsportal.u-strasbg.fr/?target=lhs1140} in the Photometric Points Section). Due to suspicion of possible contamination, we did not use 2MASS data for the effective temperature fit.}
\label{tab1140_sed}
\end{table}

\section{Discussion and Conclusions\label{sec:conclu}}

We measure a $\rm NUV_G$ luminosity for LHS~1140 of ($\rm 2.63 \pm 0.09 )\times 10^{27}\ erg\ s^{-1}$, and estimate the $\rm FUV_G$ luminosity to be $(\rm 7.97 \pm 1.57) \times 10^{26}\ erg\ s^{-1}$. 
We subsequently attempted to elucidate the FUV/NUV environment reaching the HZ planet LHS~1140b in the context of that of the present-day Earth. 
We refer to fluxes arriving outside the atmospheres, not taking into account possible atmospheric differences between Earth and LHS~1140b. Future atmospheric investigations of LHS~1140b will allow us to also put constraints on the UV environment on the planetary surface.

We estimated the present-day Earth UV flux by calculating the integrated flux of the quiet Sun in its HZ \citep{2009GeoRL..36.1101W,2012ApJ...750L..32F} in our $\rm NUV_G$ and $\rm FUV_G$ bands ($\rm NUV_{\Earth}$, $\rm FUV_{\Earth}$). We find $\rm NUV_{\Earth}=\rm 6.4 \times 10^3 \ erg \ cm^{-2} \ s^{-1}$ and $\rm FUV_{\Earth}=\rm 11.8 \ erg \ cm^{-2} \ s^{-1}$. To measure the $\rm NUV_G$ and the $\rm FUV_G$ integrated flux on LHS~1140b ($\rm NUV_{LHS}$, $\rm FUV_{LHS}$), we scaled the flux observed with {\it Swift} at the semi-major axis distance from LHS~1140 \citep[$\rm 0.0936 AU$,][]{ment} assuming that the fluxes are constant in each band.
We find $\rm NUV_{LHS}=\rm 106.8 \pm 6.2\ erg\ cm^{-2}\ s^{-1}$ ($\rm 0.017 \pm 0.001 \ NUV_{\Earth}$) and $\rm FUV_{LHS}=\rm 32.4 \pm 7.2 \ erg\ cm^{-2}\ s^{-1}$ ($\rm 2.75 \pm 0.61 \ FUV_{\Earth}$). The $\rm FUV_G/NUV_G$ integrated flux ratio reaching LHS~1140b is $\sim$100-200 times greater than the ratio reaching Earth at present. Since the GALEX UV bandpasses do not include Lyman-alpha, which dominates the FUV output of low-mass stars,  this represents a lower limit to the true top-of-atmosphere FUV/NUV ratio. 

Both FUV and NUV illumination are relevant for oxygen photochemistry. For example, FUV photons dissociating $\rm CO_2$ can be net sources of abiotic atmospheric oxygen \citep{2014E&PSL.385...22T}. Therefore, in an atmosphere with a high concentration of $\rm CO_2$ and a high FUV/NUV ratio, detection of oxygen in combination with $\rm H_2O$ does not necessarily imply the existence of life,  contrary to what was previously thought. This high FUV/NUV ratio could therefore lead to the formation of possible false-positive biosignatures: this is a warning for future atmospheric observations of LHS~1140b. As mentioned in the introduction to this paper, the photolysis of {$\rm CO_2$ is only one of the numerous abiotic oxygen-formation pathways}.

The low level of NUV flux seen by LHS~1140b could be negative for abiogenesis \citep{2017ApJ...843..110R}. This result is consistent with \citet{10.1126/sciadv.aar3302}, who place LHS~1140b out of the abiogenesis zone. In this context it is still not clear if the UV-dependent prebiotic pathways that were important for the origin of life on Earth can still proceed in a lower UV environment \citep{2017ApJ...843..110R} such as that of LHS~1140b. 
In any case, the activity evolution of M-dwarfs, even though still not completely understood, seems to decrease with age \citep{2016A&A...596A.111R, 2012ApJ...757...95C, 2007A&A...476.1373S, 2016ApJ...821...81G} implying that LHS~1140b could have suffered more high-energy irradiance in the past.
Flares could also provide the missing UV energy for abiogenesis \citep{2007Icar..192..582B}, though we do not observe flares during our observations, possibly favoring current habitability. However, we cannot exclude that flares occur in a time-span that is different from our observations or that flare activity was present in the past.

Since the relative distances between planets orbiting M-dwarfs are usually smaller than between planets around other stellar types, the probability of a panspermia scenario could increase \citep{2017PNAS..114.6689L}. 
Life could emerge on a planet with enough NUV irradiation, and eventually migrate to an outer planet where the flux of potentially destructive high-energy photons is not critical. Therefore, we estimated the $\rm NUV_G$ flux experienced by LHS~1140c \citep[semi-major axis: $\rm 0.02675 \pm 0.00070 \ AU$,][]{ment}:   $\rm 1308 \pm 43 \ erg \ cm^{-2}\ s^{-1}$, approximately 12 times greater than the $\rm NUV_G$ flux seen by LHS~1140b and 4.9 times lower than that seen by present-day Earth.

In view of the many studies of M-dwarfs considering UV radiation as deleterious for life \citep[e.g.,][]{Set, 1999OLEB...29..405H, 2007AsBio...7...30T, 2009A&ARv..17..181L, 2016PhR...663....1S, 2017AAS...22912003M} and of the UV luminosity of LHS~1140 that  we measured with {\it Swift} observations, the relatively low level and stability of UV flux experienced by LHS~1140b should be favorable for its present-day habitability.

\begin{acknowledgements} We thank the referee for useful comments and suggestions that helped to improve the paper. This study made use of data supplied by the UK {\it Swift} Science Data Centre at the University of Leicester. Part of this work is based on archival data, software, or online services provided by the Space Science Data Center-ASI. We thank B. Cenko for suggesting we look into the Pan-STARRS archive. The Pan-STARRS1 Surveys (PS1) and the PS1 public science archive have been made possible through contributions by the Institute for Astronomy, the University of Hawaii, the Pan-STARRS Project Office, the Max-Planck Society and its participating institutes, the Max Planck Institute for Astronomy, Heidelberg and the Max Planck Institute for Extraterrestrial Physics, Garching, The Johns Hopkins University, Durham University, the University of Edinburgh, the Queen's University Belfast, the Harvard-Smithsonian Center for Astrophysics, the Las Cumbres Observatory Global Telescope Network Incorporated, the National Central University of Taiwan, the Space Telescope Science Institute, the National Aeronautics and Space Administration under Grant No. NNX08AR22G issued through the Planetary Science Division of the NASA Science Mission Directorate, the National Science Foundation Grant No. AST-1238877, the University of Maryland, Eotvos Lorand University (ELTE), the Los Alamos National Laboratory, and the Gordon and Betty Moore Foundation. FB acknowledges financial support from INAF through the ASI-INAF contract 2015-019-R0.
\end{acknowledgements}

\end{document}